\documentclass[fleqn,10pt]{wlscirep}

\usepackage{amssymb}   
\usepackage{graphicx}
\usepackage{graphics}
\usepackage{mathrsfs}
\usepackage[toc,page]{appendix}
\usepackage{color}
\usepackage{fancyhdr}
\usepackage[english]{babel}
\usepackage{gensymb}
\usepackage[utf8]{inputenc}  
\usepackage[T1]{fontenc}
\usepackage{fancyvrb}
\usepackage{xcolor}
\usepackage{verbatim}
\usepackage{color}
\usepackage{amsmath}

\newcommand{\ie}{\emph{i.e.}$\;$}

\newcommand{\vect}[1]{\boldsymbol{#1}}

\title{Improving landscape inference by integrating heterogeneous data in the inverse Ising problem}

\author[a,+]{Pierre Barrat-Charlaix}
\author[a,b,+]{Matteo Figliuzzi} 
\author[a,*]{Martin Weigt}

\affil[a]{Sorbonne Universit\'es, UPMC Univ Paris 06, CNRS, Biologie computationnelle et quantitative - Institut de Biologie Paris Seine, 75005 Paris, France}
\affil[b]{Sorbonne Universit\'es, UPMC Univ Paris 06, Institut de Calcul et de la Simulation, 75005 Paris, France}
\affil[+]{these authors contributed equally to this work}
\affil[*]{To whom correspondence should be addressed. Email: martin.weigt@upmc.fr}

\begin{abstract}
The inverse Ising problem and its generalizations to Potts and continuous spin models have recently attracted much attention thanks to their successful applications in the statistical modeling of biological data. In the standard setting, the parameters of an Ising model (couplings and fields) are inferred using a sample of equilibrium configurations drawn from the Boltzmann distribution. However, in the context of biological applications, quantitative information for a limited number of microscopic spins configurations has recently become available. In this paper, we extend the usual setting of the inverse Ising model by developing an integrative approach combining the equilibrium sample with (possibly noisy) measurements of the energy performed for a number of arbitrary configurations. Using simulated data, we show that our integrative approach outperforms standard inference based only on the equilibrium sample or the energy measurements, including error correction of noisy energy measurements. As a biological proof-of-concept application, we show that mutational fitness landscapes in proteins can be better described when combining evolutionary sequence data with complementary structural information about mutant sequences. 
\end{abstract}
\begin{document}

\flushbottom
\maketitle
%
%
\thispagestyle{empty}

\section*{Introduction}

High-dimensional data characterizing the collective behavior of complex systems are increasingly available across disciplines. A global statistical description is needed to unveil the organizing principles ruling such systems and to extract information from raw data. Statistical physics provides a powerful framework to do so. A paradigmatic example is represented by the Ising model and its generalizations to Potts and continuous spin variables, which have recently become popular for extracting information from large-scale biological datasets. Successful examples are as different as multiple-sequence alignments of evolutionary related proteins \cite{weigt2009identification,mora2010maximum,ferguson2013translating}, gene-expression profiles \cite{lezon2006using}, spiking patterns of neural networks \cite{schneidman2006weak,cocco2009neuronal}, or the collective behavior of bird flocks \cite{bialek2012statistical}. This widespread use is motivated by the observation that the least constrained (i.e. maximum-entropy \cite{jaynes1957information} ) statistical model reproducing empirical single-variable and pairwise frequencies observed in a list of equilibrium configurations is given by a Boltzmann distribution:
\begin{equation}\label{eq:Hamiltonian_def}
	P(\vect{s})=\frac{1}{\mathcal{Z}}\exp\left\{-\mathcal{H}(\vect{s})\right\}, \quad \mathcal{H}=-\sum_{i<j}^N J_{ij}s_i s_j - \sum_{i=1}^N h_i s_i\ ,
\end{equation}
with $\vect{s}=(s_1,...,s_N)$ being a configuration of $N$ binary variables or 'spins'. Inferring the couplings $\mathbf{J} =\{J_{ij}\}_{1\leq i<j\leq N}$  and fields $\mathbf{h}=\{h_i\}_{1\leq i\leq N}$ in the Hamiltonian $\mathcal{H}$ from data, known as the {\em inverse Ising problem}, is computationally hard for large systems ($N\gg 1$). It involves the calculation of the partition function $\mathcal{Z}=\sum_{\vect{s}}e^{-\mathcal{H}(\vect{s})}$ as a sum over an exponential number of configurations. The need to develop efficient  approximate approaches has recently triggered important work within the statistical-physics community, cf. e.g. \cite{roudi2009ising,sessak2009small,mezard2009constraint,cocco2011high,cocco2011adaptive,nguyen2012mean,aurell2012inverse,nguyen2012bethe,decelle2014pseudolikelihood}.

Despite the broad interest in inverse problems, the methodological setting has remained rather limited: all of this literature, including the biological cases mentioned in the beginning, seeks to estimate model parameters starting from a set of configurations $\vect{s}$, which are considered to be at equilibrium and independently drawn from $P(\vect{s})$. 
Real data, however, may be quite different. In biological systems, ``microscopic spins configurations'' (e.g. amino-acid sequences) are increasingly accessible to experimental techniques, and {\it quantitative information} for a limited number of  {\it particular} configurations (e.g. three-dimensional structures, measured activities or thermodynamic stabilities for selected proteins) is frequently available. It seems reasonable to actually integrate such information into the inverse Ising problem instead of ignoring it. In this work, we use two different types of data (cf.~Fig.~\ref{fig:Scheme}):
\begin{itemize}
\item As in the standard inverse Ising problem, part of the data comes as a sample of equilibrium configurations assumed to be drawn from the Boltzmann distribution to be inferred.
\item The second data source is a collection of arbitrary configurations together with {\it noisy} measurements of their energy.
\end{itemize}
These data sets are limited in size and accuracy. Therefore an optimized integration of both data types is expected to improve the overall performance as compared to the individual use of one single data set.

The inspiration to develop this new integrative framework for the inverse Ising problem is taken from {\it protein fitness landscapes} in biology, which provide a quantitative mapping from any amino-acid sequence $\vect{s}=(s_1,...,s_N)$ to a fitness $\phi(\vect{s})$ measuring the ability of the corresponding protein to perform its biological function. Fitness landscapes are of outstanding importance in evolutionary and medical biology, but it appears impossible to deduce a protein's fitness from its sequence only. Experimental or computational approaches exploiting other data are urgently needed.

Information about fitness landscapes can be found in the amino-acid statistics observed in natural protein sequences, which are related to the protein of interest. In fact they represent diverse but functional configurations sampled by evolution. It has been recently proposed that their statistical variability can be captured by Potts models (generalization of the Ising model to 21-state amino-acid variables).  Indeed, statistical models inferred from large collections of natural sequences have recently led to good predictions of experimentally measured effects \cite{figliuzzi2016coevolutionary,asti2016maximum,morcos2014coevolutionary,chakraborty2014hiv}: in a number of systems, the fitness cost $\Delta \phi(\vect{s})\equiv \phi(\vect{s})-\phi(\vect{s}^{\rm ref})$ of mutating any amino acid in a reference protein $\vect{s}^{\rm ref}$ strongly correlates with the corresponding energy changes in the inferred statistical model,
\begin{eqnarray}
\Delta \phi(\vect{s}) \sim & log \left[\frac{P(\vect{s})}{P(\vect{s}^{\rm ref})}\right] = -\big(\mathcal{H}(\vect{s})-\mathcal{H}(\vect{s}^{\rm ref})\big),
\end{eqnarray} 
suggesting that the Hamiltonian of the inferred models is strictly related to the underlying mutational landscapes. 

While evolutionary diverged sequences can be regarded as a global sample of the fitness landscape, further information can be obtained from direct measurements on particular {\it 'microstates'} of the  system, \emph{i.e.} individual protein sequences. Recent advances in experimental technology allow for conducting large-scale {\it mutagenesis} studies: in a typical experiment, a reference protein of interest is chosen, and a large number ($10^3-10^5$) of mutant proteins (having sequences differing by one or few amino acids from the reference) are synthesized and then characterized in terms of fitness. This provides a systematic {\it local} measurement of the fitness landscape \cite{mclaughlin2012spatial,jacquier2013capturing,melamed2013deep}.
Regression analysis may be used to globally model mutational landscapes \cite{hinkley2011systems}.
A second-order parameterization of $\phi$ arises naturally in this context, when considering an expansion of effects in terms of independent additive effects and pairwise {\it 'epistatic'} interactions between sites \cite{de2014empirical},
\begin{equation}
\phi(\vect{s})=\sum_{i=1}^N \varphi_i(s_i)+ \sum_{1\leq i\leq j\leq N} \varphi_{i,j}(s_i,s_j)\ .
\end{equation}
However, the number of accessible mutant sequences remains small compared to the number of terms in this sum, and mutagenesis data alone are not sufficient to faithfully model fitness landscapes \cite{Otwinowski03062014}. 

In situations where no single dataset is sufficient for accurate inference, integrative methods accounting for complementary data sources will improve the accuracy of computational predictions. In this paper, (i) we define a generalized inference framework based on the availability of an equilibrium sample {\it and} of complementary quantitative information; (ii) we propose a Bayesian integrative approach to improve over the limited accuracy obtainable using standard inverse problems; (iii) we demonstrate the practical applicability of our method in the context of predicting mutational effects in proteins, a problem of outstanding bio-medical importance for questions related to genetic disease and antibiotic drug resistance.\\

\section*{Results}
\subsection*{An integrated modeling} 
\label{sec:an_integrated_modelling}

\begin{figure}[htb]
         \centering
		 \includegraphics[width=0.85\textwidth]{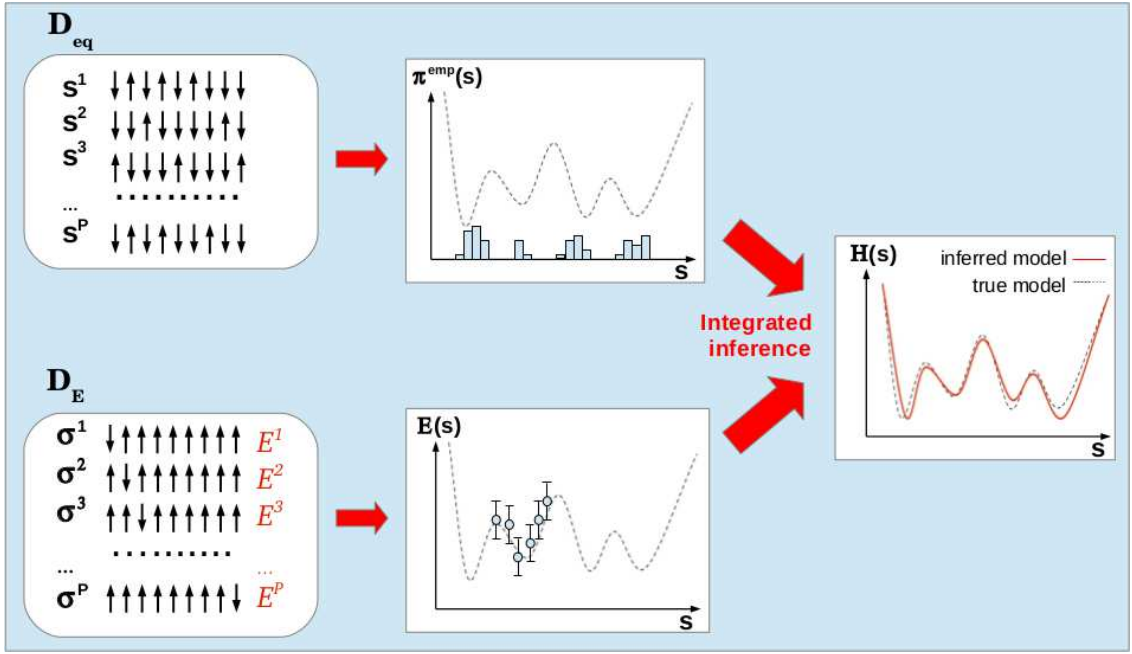}
         \caption{Schematic representation of the inference framework: a sample of equilibrium configurations (dataset  $\mathbf{D}_{\rm eq}$) and noisy energy measurements for another set of configurations (dataset  $\mathbf{D}_{\rm E}$) are integrated within a Bayesian approach to infer the model $\mathcal{H}$. The dashed lines represent the underlying true landscape, which has to be inferred, the red line the inferred landscape. 
}
\label{fig:Scheme}
 \end{figure}

\textbf{The inference setting.} Inspired by this discussion, we consider two different datasets originating from a \emph{true} model $\mathcal{H}^{0}$. The first one, ${\mathbf D}_{\rm eq} = \{ \vect{s}^1,...,\vect{s}^M \}$, is a collection of $M$ equilibrium configurations independently drawn from the Boltzmann distribution $P^{0}(\vect{s})$.
For simplicity, we consider binary variables $s_i\in\{0,1\}$, corresponding to a "lattice-gas" representation of the Ising model in Eq.~(\ref{eq:Hamiltonian_def}). This implies that energies are measured with respect to the reference configuration $\vect{s}^{\rm ref}=(0,...,0)$. The standard approach to the inverse Ising problem uses only this type of data to infer parameters of $\mathcal{H}$: couplings and fields in Eq.~(\ref{eq:Hamiltonian_def}) are fitted so that the inferred model reproduces the empirical single and pairwise frequencies, 
\begin{equation}\label{eq:Observables_def}
	\pi_i^{\text{emp}} = \frac{1}{M}\sum_{\mu=1}^M s_i^{\mu}, \qquad \pi_{ij}^{\text{emp}} = \frac{1}{M}\sum_{\mu=1}^M s_i^{\mu}s_j^{\mu}\ .
\end{equation}

The second dataset provides a complementary source of information, which shall be modeled as noisy measurements of the {\it energies} of a set of $P$ {\it arbitrary} (i.e. not necessarily equilibrium) configurations $\vect{\sigma}^a$. These data are collected in the dataset  $\mathbf{D}_E = \{(E^a, \vect{\sigma}^a)_{a=1,...,P}\}$, with
\begin{equation}\label{eq:LocalData_def}
	 E^a = \mathcal{H}^{0}(\vect{\sigma}^a) + \xi^a \quad a=1,...,P\ .
\end{equation}
The noise $\xi^a$ models measurement errors or uncertainties in mapping measured quantities to energies of the Ising model. For simplicity,  we consider $\xi^a$ to be white Gaussian noise with zero mean and variance $\Delta^2$: $\langle \xi^a \xi^b \rangle=\delta_{a,b}\Delta^2$.

As schematically represented in Fig.~\ref{fig:Scheme}, datasets $\mathbf{D}_{\rm eq}$ and $\mathbf{D}_E$ constitute different sources of information about the energy landscape defined by Hamiltonian $\mathcal{H}^{0}$. Observables in Eq.~(\ref{eq:Observables_def}) are empirical averages computed from equilibrium configurations in $\mathbf{D}_{\rm eq}$, providing \emph{global} information about the energy landscape. On the contrary, configurations in $\mathbf{D}_E$ are arbitrarily given, and a (noisy) measurement of their energies provides \emph{local} information on particular points in the landscape. \\

\textbf{A maximum-likelihood approach.} To infer the integrated model, we consider a {\it joint description of the probabilities of the two data types} for given parameters $\mathbf{J}$ and $\mathbf{h}$ of tentative Hamiltonian $\mathcal{H}$. The probability of observing the sampled configurations in $\mathbf{D}_{\rm eq}$ equals the product of the Boltzmann probability of each configuration, 
\begin{equation}\label{eq:P_DMC_def}
	P(\mathbf{D}_{\text{eq}}\vert \mathbf{J},\mathbf{h}) =  \exp\left\{-\sum_{\mu=1}^P \mathcal{H}(\mathbf{s}^{\mu}) - M\log\mathcal{Z}(\mathbf{J},\mathbf{h})\right\}\ .
\end{equation}
To derive an analogous expression for the second dataset, we integrate over the Gaussian distribution of the noise $\xi^a = E^a-\mathcal{H}(\vect{\sigma}^a)$ obtaining a Gaussian probability of the energies (remember configurations in $\mathbf{D}_E$ are arbitrarily given):
\begin{eqnarray}\label{eq:P_DE_def}
	P(\mathbf{D}_E\vert\mathbf{J},\mathbf{h})  \!\!\!\!\! &=&  \!\!\!\!\! \prod_{a=1}^P\int  d\xi_a P(\xi_a)\ \delta( E^a- \mathcal{H}(\vect{\sigma}^a) - \xi^a) \nonumber\\
	\! &=& \! \!\!\!\! \frac{1}{(2\pi\Delta^2)^{\frac P2}}\exp\left\{\! -\sum_{a=1}^P\! \frac{\left[ E^a-\mathcal{H}(\vect{\sigma}^a)\right]^2}{2\Delta^2}\! \right\}
\end{eqnarray}
The combination of these expressions provides the joint log-likelihood for the model parameters given the data: 
\begin{equation}\label{eq:JointLikelihood_def}
\mathcal{L}(\mathbf{J},\mathbf{h}\vert\mathbf{D}_{\text{eq}},\mathbf{D}_E) = \log P(\mathbf{D}_{\text{eq}}\vert \mathbf{J},\mathbf{h}) +\log P(\mathbf{D}_E\vert\mathbf{J},\mathbf{h})
\end{equation}
Maximizing the above likelihood with respect to parameters $\{h_i\}_{1\leq i\leq N}$ and $\{J_{ij}\}_{1\leq i<j\leq N}$ leads to the following self-consistency equations:
\begin{eqnarray}\label{eq:MaxLikelihood_lambda}
p_i(\mathbf{J},\mathbf{h}) &=& \pi_i^{\text{emp}} + \frac{\lambda}{1-\lambda}\frac{1}{M}\sum_{a=1}^P\sigma_i^a\left[E^a - \mathcal{H}(\vect{\sigma}^a)\right] \\
p_{ij}(\mathbf{J},\mathbf{h}) &=& \pi_{ij}^{\text{emp}} + \frac{\lambda}{1-\lambda}\frac{1}{M}\sum_{a=1}^P\sigma_i^a\sigma_j^a\left[E^a - \mathcal{H}(\vect{\sigma}^a)\right] \nonumber
\end{eqnarray} 
with $p_i(\mathbf{J},\mathbf{h})=\langle \sigma_i \rangle_{\mathcal{H}}$, $p_{ij}(\mathbf{J},\mathbf{h})=\langle \sigma_i \sigma_j \rangle_{\mathcal{H}}$ being single and pairwise averages in the model Eq.~(\ref{eq:Hamiltonian_def}). We have introduced the parameter $\lambda=\frac{1}{1+\Delta^2}$: in practical applications, the error $\Delta$ may not be known, and the parameter $0\leq \lambda<1$ allows to weigh data sources differently. For $\lambda=0$ (\ie large noise),  the standard inverse Ising problem is recovered: optimal parameters are such that the model exactly reproduces magnetizations and correlations of the sample. For $\lambda>0$, the second dataset containing quantitative data is taken into account: whenever energies computed from the Hamiltonian $\mathcal{H}$ do not match the measured ones, the model statistics deviates from the sample statistics. 
Both log-likelihood terms in (\ref{eq:JointLikelihood_def}) are concave, and thus their sum: Eq. (\ref{eq:MaxLikelihood_lambda}) has a unique solution.\\

\textbf{Noiseless measurements.} The case of noiseless energy measurements in Eq.~(\ref{eq:LocalData_def}) (i.e. $\lambda\to 1$) has to be treated separately. First, energies have to be perfectly fitted by the model, by solving the following linear problem:
\begin{equation}\label{eq:linear_sys}
\mathbf{X}\vec{\mathcal{H}} =\vec{E}\ ,
\end{equation}
where we have  introduced a $N(N+1)/2$ dimensional vectorial representation $\vec{\mathcal{H}}=(h_1,...,h_N,J_{12},...,J_{N,N-1})^T$ of the model parameters, and $\vec{E}=(E_1,...,E_P)^T$ contains the exactly measured energies. The  matrix
\begin{equation}
\mathbf{X}=\begin{bmatrix}
\sigma_1^1 & \cdots  & \sigma_N^1 & (\sigma_1^1\sigma_2^1) & \cdots & (\sigma_{N-1}^1 \sigma_N^1) \\
\vdots & & \vdots &\vdots & & \vdots \\
\sigma_1^P &\cdots& \sigma_N^P & (\sigma_1^P\sigma_2^P) &\cdots & (\sigma_{N-1}^P \sigma_N^P) \\
\end{bmatrix}
\end{equation} 
specifies which parameters contribute to the energies of configurations in the second dataset. If $K = N(N+1)/2-rank(X) >0$, the parameters cannot be uniquely determined from the measurements: The sample $\mathbf{D}_{\rm eq}$ can be used to remove the resulting degeneracy. To do so, we parametrize the set of solutions of Eq.~(\ref{eq:linear_sys}) as follows:
\begin{equation}
\vec{\mathcal{H}}= \vec{\mathcal{H}}_{nh} +\sum_{k=1}^K \alpha_k \vec{\mathcal{O}}_k
\end{equation}
where $\vec{\mathcal{H}}_{nh}$ is any particular solution of the non homogeneous Eq.~(\ref{eq:linear_sys}), and $\{\vec{\mathcal{O}}_k\}$ a basis of observables spanning the null space of the associated homogeneous problem $\mathbf{X}\vec{\mathcal{H}} =0$. The free parameters $\alpha_k \in \mathbb{R}$ can be fixed by maximizing their likelihood given sample $\mathbf{D}_{\rm eq}$,
\begin{equation}\label{eq:alpha}
	\mathcal{L}(\alpha | \mathbf{D}_{\rm eq} ) \propto\exp\left\{-\sum_{k,\mu} \alpha_k \mathcal{O}_k(\mathbf{s^{\mu}}) - M\log\mathcal{Z}(\mathcal{\alpha})\right\}
\end{equation}
with $\mathcal{O}_k(\mathbf{s})=(s_1, \cdots, s_N, (s_1s_2), \cdots, (s_{N-1} s_N)) \cdot \vec{\mathcal{O}}_k$.
The maximization provides conditions for the observables ($k=1,...,K$),
\begin{equation}\label{eq:beta_max}
	\frac{\partial\mathcal{L}(\mathbf{\alpha})}{\partial \alpha^k} \propto \frac{1}{M}\sum_{\mu=1}^M\mathcal{O}_k(\mathbf{s^\mu})-\langle \mathcal{O}_k\rangle_\mathcal{H} = 0 \ .
\end{equation} 
Eq.~(\ref{eq:beta_max}) shows that the $\alpha_k$ have to be fixed such that empirical averages $\frac{1}{M}\sum_{\mu=1}^M\mathcal{O}_k(\mathbf{s^\mu})$ equal model averages $\langle \mathcal{O}_k\rangle_\mathcal{H}$.
Any possible sparsity of the matrix of measured configurations $\mathbf{X}$ (entries are 0 or 1 by definition) can be exploited to find a sparse representation of the $\{\vec{\mathcal{O}}_k\}$. In the protein example discussed above, mutagenesis experiments typically quantify all possible single-residue mutations of a reference sequence (denoted $(0,...,0)$ without loss of generality). In this case, the pairwise quantities $s_is_j$ with $1<i<j<N$ can be chosen as the basis $\{\mathcal{O}_k\}$ of the null space. A particular solution of the non-homogeneous system (\ref{eq:linear_sys}) is given by the paramagnetic Hamiltonian $\mathcal{H}_{nh}=\sum_i E^is_i$, with $E^i$ being the energy shift due to spin flip $s_i=0\mapsto 1$.




\subsection*{Artificial data} 
\label{sub:artificial_data}

We first evaluate our method on artificial data (\emph{Materials and Methods}). Random couplings $\mathbf{J}^{0}$ and fields $\mathbf{h}^{0}$ are chosen for a system of $N=32$ spins. Dataset $\mathbf{D}_{\text{eq}}$ is created by Markov chain Monte Carlo (MCMC) sampling, resulting in $M=100$ equilibrium configurations. To mimic a protein 'mutagenesis' experiment, one of these configurations is chosen at random as the reference sequence, and the energies of all $N$ configurations differing by a single spin flip from the reference (thereafter referred to as single mutants) are calculated, resulting in dataset $\mathbf{D}_E$ (after adding noise of standard deviation $\Delta_0$). Datasets $\mathbf{D}_E$ and $\mathbf{D}_{\text{eq}}$ will subsequently called ``local'' and ``global'' data respectively.

Equations~(\ref{eq:MaxLikelihood_lambda}) are solved using steepest ascent, updating parameters $\mathbf{J}$ and $\mathbf{h}$ in direction of the gradient of the joint log-likelihood (\ref{eq:JointLikelihood_def}). Since the noise $\Delta_0$ may not be known in practical applications, we solve the equations for several values of $\lambda\in[0,1]$, weighing data sources differently. We expect the optimal inference to take place at a value $\lambda$ that maximizes the likelihood in Eq.~(\ref{eq:JointLikelihood_def}), \ie $\lambda_0=(1+\Delta_0^2)^{-1}$. For $\lambda = 0$, this procedure is equivalent to the classical Boltzmann machine \cite{ackley1985learning}, but for $\lambda>0$, the term corresponding to the quantitative essay constrains energies of sequences in $\mathbf{D}_E$ to stay close to the measurements. As explained above, the case $\lambda = 1$ has to be treated separately; a similar gradient ascent method is used. Since exact calculations of gradients are computationally hard, the mean-field approximation is used (\emph{Materials and Methods}).

To evaluate the accuracy of the inference, most of the existing literature on the inverse Ising modeling simply compares the inferred parameters with the true ones. However, a low error in the estimation of each inferred parameter does not guarantee that the inferred distribution matches the true one. On the contrary, in the case of a high susceptibility of the statistics with respect to parameter variations, or if the estimation of parameters is biased, the distributions of the inferred and true models could be very different even for small errors on individual parameters. For this reason, we introduce two novel evaluation procedures. First, to estimate the accuracy of the model on a local region of the configuration space, we test its ability to reproduce \emph{energies} of configurations in $\mathbf{D}_E$. Then, we estimate the global similarity of true and inferred distributions using a measure from information theory. \\

\begin{figure}[h!]
         \centering
		 \includegraphics[width=0.85\textwidth]{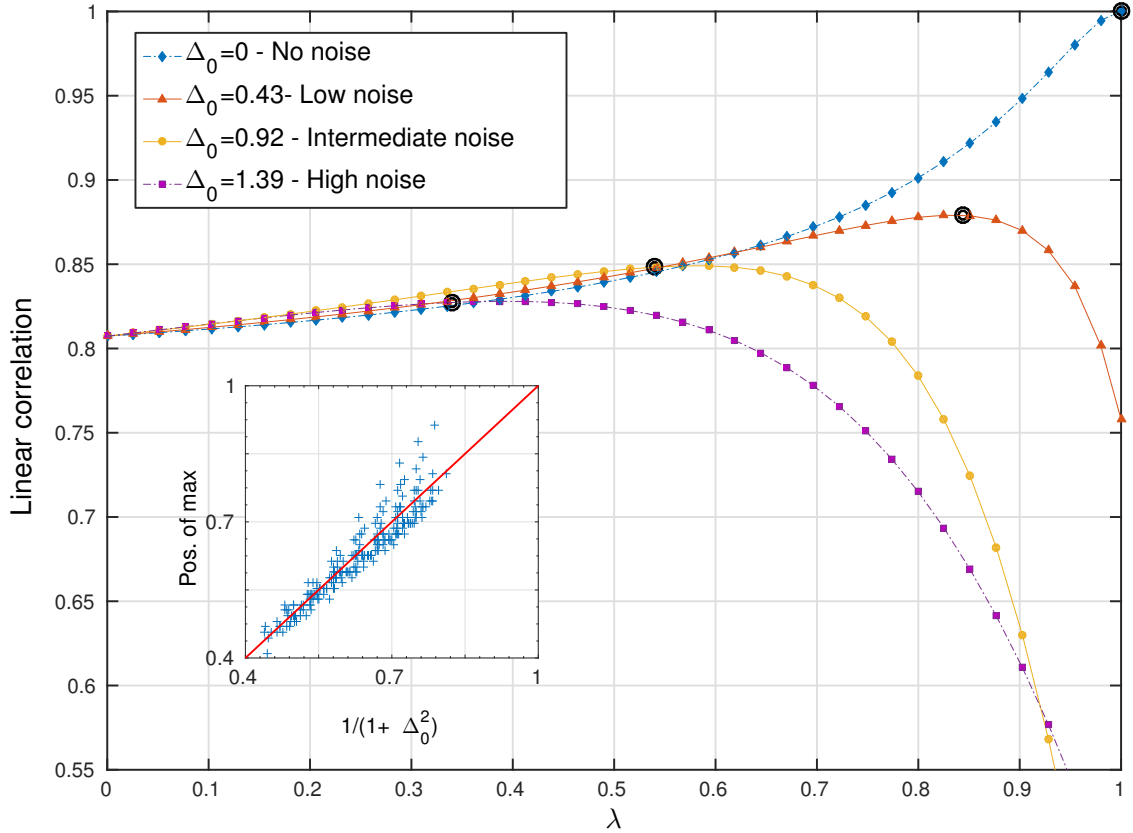}
         \caption{
Integration of noisy measurements of energies of single mutants with $M$ equilibrium configurations. The linear correlation between predicted and true single mutants energies is shown in dependence of integration parameter $\lambda$, for four different values of the noise $\Delta_0 = 0,0.43,0.92,1.39$ added to the energies in the second data set. The integration strength $\lambda_0 = 1/(1+\Delta_0^2)$, which would be naturally used in case of an {\it a priori} known noise level, is located close to the optimal inference, cf. the black circles for $\lambda_0 = 1,0.84,0.54,0.34$.  The \textbf{insert} shows the value of the integration strength $\lambda$ reaching maximal correlation, as a function of the theoretical value  $1/(1+\Delta_0^2)$, for 200 independent realisations of the input data at different noise levels. Points are found to be closely distributed around the diagonal (red line).
}
\label{fig:SingleMutants_prediction}
 \end{figure}

 
\textbf{Error correction of local data.} We first test the ability of our approach to predict the true single-mutant energies, when noisy measurements are presented in $\mathbf{D}_E$, \ie to correct the measurement noise using the equilibrium sample $\mathbf{D}_{\rm eq}$. For every $\lambda$, $\mathbf{J}$ and $\mathbf{h}$ are inferred and used to compute predicted energies of the $N$ configurations in $\mathbf{D}_E$. The linear correlation between such predicted energies (measured with the inferred Hamiltonian) and the true energies (measured with the true Hamiltonian) is plotted as a function of $\lambda$ in Fig.~\ref{fig:SingleMutants_prediction}.

In the very low noise regime, $\Delta_0\simeq 0$, the top curve in Figure \ref{fig:SingleMutants_prediction} reaches its peak at $\lambda\simeq 1$, which is expected as local data is then sufficient to accurately ``predict'' energies from single mutants. On the contrary, in the high noise regime, the maximum is located close to $\lambda= 0$, pointing to the fact that local data is of little use in this case. Between those two extremes, an optimal integration strength can be found, yielding a better prediction of energies in $\mathbf{D}_{\text{E}}$ as for any of the datasets taken individually. It is interesting to notice that even for highly noisy data, integrating the two sources of information with the right weight $\lambda$ results in an improved modeling. 

The insert of Fig.~\ref{fig:SingleMutants_prediction} shows the integration strength $\lambda$ at which the best correlation is reached, against the corresponding theoretical value $\lambda_0=1/(1+\Delta_0^2)$ for different realizations. On average, optimal integration is reached close to the theoretical case of equation (\ref{eq:JointLikelihood_def}). This result highlights the possibility of using this integrative approach to \emph{correct measurement errors} in the energies of single mutants. If a dataset such as $\mathbf{D}_{\text{eq}}$ provides global information about the energy landscape, and the measurement noise $\Delta_0$ can be estimated, an appropriate integration can then be used to infer more accurately the energies of the single mutants. \\

 \begin{figure}[h!]
          \centering
          \includegraphics[width=0.85\textwidth]{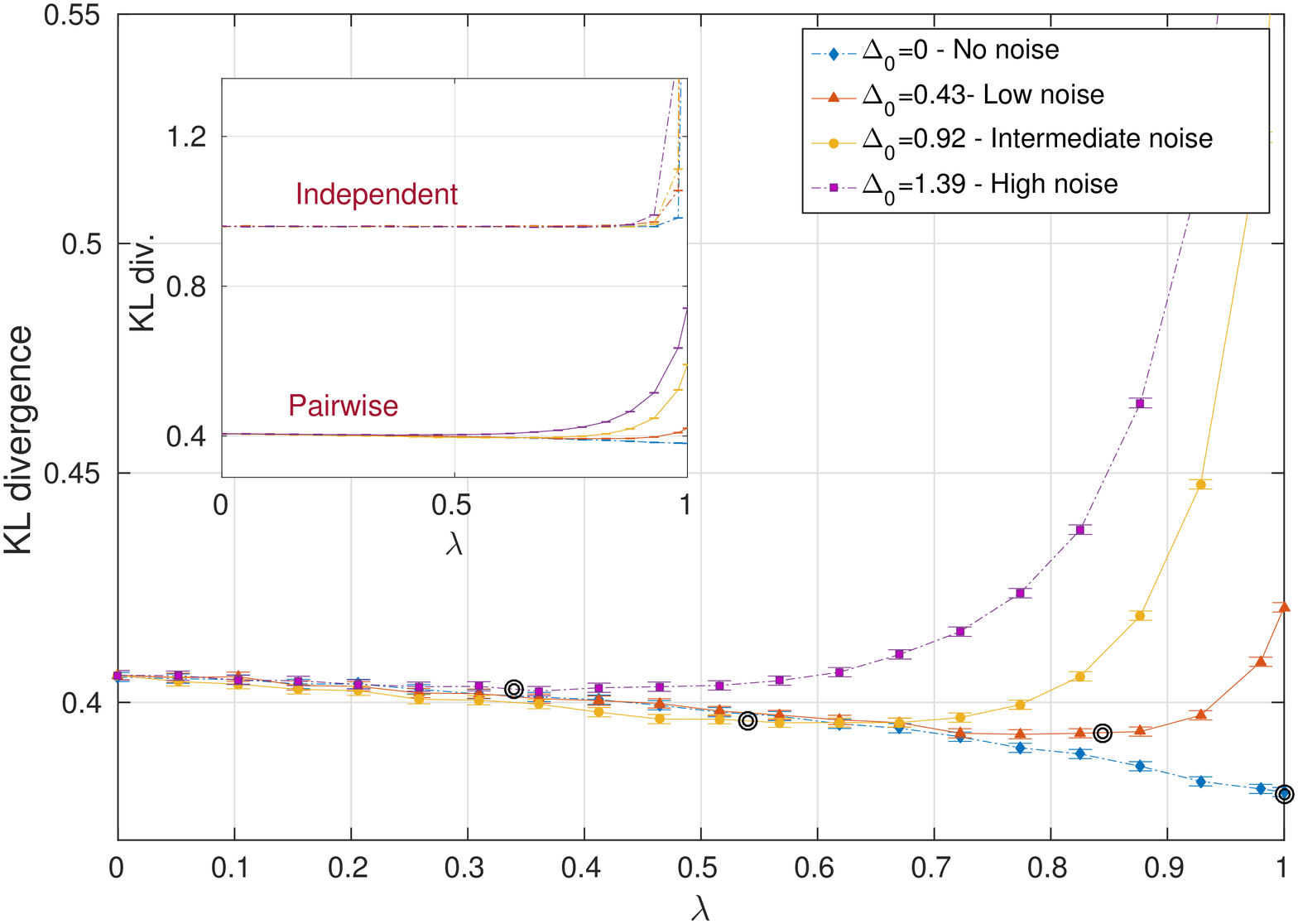}
          \caption{Symmetric Kullback-Leibler divergence $\Sigma$ between true and integrated pairwise models versus strength of integration $\lambda$, estimated from $M_K=3\cdot 10^6$ MCMC samples. Different curves correspond to different noise levels added to single mutants energies used for integration, so that datasets $\mathbf{D}_E$ are the same as in Fig.~\ref{fig:SingleMutants_prediction}. \textbf{Insert -} Comparison with an independent model using only fields $\mathbf{h}$, with the same methodology. }
 \label{fig:DoubleMutants_prediction}
  \end{figure}
 
\textbf{Global evaluation of the inferred Ising model.} To assess the ability of our integrative procedure to provide a {\it globally} accurate model, we use the Kullback-Leibler divergence $D_{\scriptscriptstyle KL}(P^{0}||P)$ between the true model $P^{0}\propto e^{-{\mathcal{H}^{0}}}$ and the inferred $P\propto e^{-\cal H}$ (\emph{Materials and Methods}). The symmetric expression
\begin{eqnarray}
\label{eq:kl}
 \Sigma(P^{0},P) &=&\; D_{\scriptscriptstyle KL}(P^{0}||P) + D_{\scriptscriptstyle KL}(P||P^{0}) \nonumber\\
&=& \langle \mathcal{H}^{0}-\mathcal{H}\rangle_{P} + \langle \mathcal{H}-\mathcal{H}^{0}\rangle_{{P}^{0}} 
\end{eqnarray}
simplifies to the average difference between true and inferred energies. It can be consistently estimated using MCMC samples ${\mathbf D}$ (resp.~${\mathbf D}^{0}$) drawn from $P$ (resp.~$P^{0}$), without the need to calculate the partition functions. $\Sigma(P^{0},P)$ has an intuitive interpretation in terms of distinguishability of models: It represents the log-odds ratio between the probability to observe samples  ${\mathbf D}$ and ${\mathbf D}^{0}$ in their respective generating models, and the corresponding probability with models $\mathcal{H}$ and $\mathcal{H}^{0}$ swapped:
\begin{equation}
\Sigma(P^{0},P)= \frac{1}{M_{K}}\log \left[ \frac{P({\mathbf D}|\mathcal{H})P({\mathbf D}^{0}|\mathcal{H}^{0})}{P({\mathbf D}|\mathcal{H}^{0})P({\mathbf D}^{0}|\mathcal{H})} \right] \ .
\end{equation}
where $M_K$ is the number of sampled configurations in ${\mathbf D}^{\mathcal{H}}$ and ${\mathbf D}^{0}$.\\
The inferred model undoubtedly benefits from the integration, as a minimal divergence between the generating and the
inferred probability distributions is found for an intermediate value of $\lambda$,
outperforming both datasets taken individually (Figure \ref{fig:DoubleMutants_prediction}). It has to be noted that even in the noiseless case 
$\Delta_0 = 0$, the minimum in $\Sigma(P^{0},P)$ obtained at $\lambda = 1$ depends crucially on the availability of the equilibrium sample
$D_{\rm eq}$. The local data $D_E$ are not sufficient to fix uniquely all model parameters, and the degeneracy in parametrization is resolved using
$D_{\rm eq}$ as explained at the end of Sec.~\ref{sec:an_integrated_modelling}.

As a comparison, the same analysis is done using an independent modeling that uses only fields $\vect{h}$, and no couplings. The inset of Figure \ref{fig:DoubleMutants_prediction} clearly shows that the pairwise modeling outperforms the independent one. Even the limit $\lambda\to 1$, where 
$D_{\rm eq}$ becomes irrelevant in the independent model, the performance of the integrative pairwise scheme is not attained.\\


\subsection*{Biological data} 
\label{sub:biological_data}

To demonstrate the practical utility of our integrative framework, we apply it to the challenging problem of predicting the effect of amino-acid mutations in proteins. 
To do so, we use three types of data: $(i)$ Multiple-sequence alignments (MSA) of homologous proteins containing large collections of sequences with shared evolutionary ancestry and conserved structure and function; they are obtained using HMMer \cite{mistry2013challenges} using profile models from the Pfam database \cite{finn2013pfam}. Due to their considerable sequence divergence (typical Hamming distance $\sim 0.8 N$), they provide a \textit{global} sampling of the underlying fitness landscape. $(ii)$ Computational predictions of the impact of all single amino-acid mutations on a protein's structural stability \cite{dehouck2011popmusic} are used to \textit{locally} characterize the fitness landscape around a given protein. The noise term $\xi^a$ represents the limited accuracy of this predictor, and the uncertainty in using structural stability as a proxy of protein fitness. $(iii)$ Mutagenesis experiments have been used before to simultaneously quantify the fitness effects of thousands of mutants \cite{jacquier2013capturing,mclaughlin2012spatial}. While datasets $(i)$ and $(ii)$ play the role of $\mathbf{D}_{eq}$ and $\mathbf{D}_E$ in inference, dataset $(iii)$ is used to assess the quality of our predictions (ideally one would use the most informative datasets (i) and (iii) to have maximally accurate predictions, but no complementary dataset to test predictions would be available in that case).

To apply the inference scheme to such protein data, three modifications with respect to simulated data are needed. First, the relevant description in this case is a 21-state Potts Model (\emph{Supporting Information}), since each variable $s_i, i=1,...,N,$ can now assume 21 states (20 amino acids, one alignment gap) \cite{morcos2011direct}. Second, since measured fitnesses and model energies are found in a monotonous non-linear relation, we have used the robust mapping introduced in \cite{figliuzzi2016coevolutionary} (reviewed in the \emph{Supporting Information}). Third,  since correlations observed in MSA are typically too strong for the MF approximation to accurately estimate marginals, we relied on Markov Chain Monte Carlo (\emph{Materials and Methods}), which has recently been shown to outperform other methods in accuracy of inference for protein-sequence data \cite{sutto2015pnas,haldane2016structural}.

 \begin{figure}[h!]   
                 \includegraphics[width=0.85\textwidth]{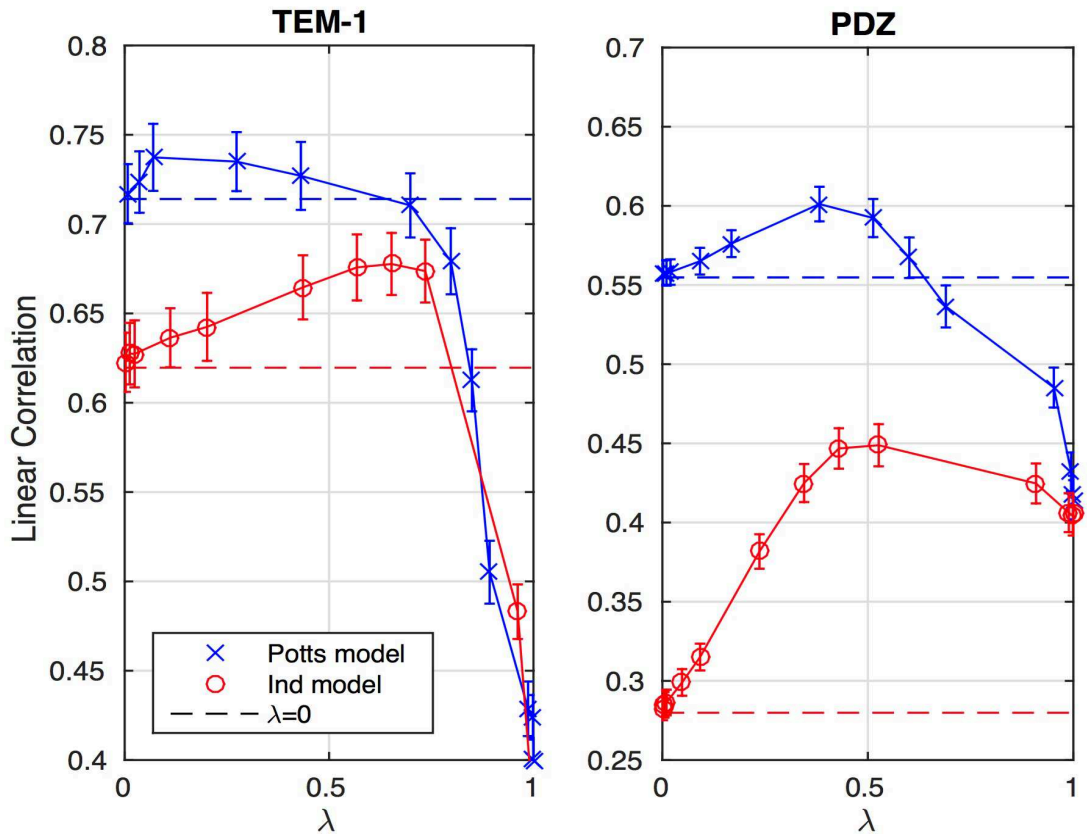}
    \caption{\label{fig:protein}Linear correlation between experimental fitness effects and predictions from integrated models, at different values of  $\lambda$, for 742 single mutations in the beta-lactamase TEM-1 (left panel) and for 1426 single mutations in the PSD-95 PDZ domain (right panel). Error bars represent statistical errors estimated via jack-knife estimation.
          }
\label{fig:SingleMutants_realdata}
 \end{figure}

We have tested our approach for predicting the effect of single amino-acid mutations in two different proteins: the $\beta$-lactamase TEM-1, a bacterial enzyme providing antibiotic resistance, and the PSD-95 signaling domain belonging to the PDZ family. In both systems computational predictions can be tested against recent high-throughput experiments quantifying the \textit{in-vivo} functionality of thousands of protein variants \cite{jacquier2013capturing,mclaughlin2012spatial}.
Fig.~\ref{fig:protein} shows the Pearson correlations between inferred energies and measured fitnesses as a function of the weight $\lambda$: Maximal accuracy is achieved at finite values of $\lambda$ when both sources of information are combined, significantly increasing the predictive power of the models inferred considering the statistics of homologs only ($\lambda=0$).
When repeating the integrated modeling with a paramagnetic model where all sites are treated independently, $\mathcal{H}^{\rm ind}(\mathbf{a})=-\sum_{i=1}^N h_i(a_i)$ (only single-site frequencies are fitted in this case) the predictive power drops as compared to the Potts model, cf.~the red lines in Fig.~\ref{fig:protein}.



\section*{Conclusion}

In this paper, we have introduced an integrative Bayesian framework for the inverse Ising problem. In difference to the standard setting, which uses only a global sample of independent equilibrium configurations to reconstruct the Hamiltonian of an Ising model, we also consider a local quantification of the energy function around a reference configuration. Using simulated data, we show that the integrated approach outperforms inference based on each single dataset alone. The gain over the standard setting of the inverse Ising problem is particularly large when the equilibrium sample is too small to allow for accurate inference. 

This undersampled situation is particularly important in the context of biological data. The prediction of mutational effects in proteins is of enormous importance in various bio-medical applications, as it could help understanding complex and multifactorial genetic diseases, the onset and the proliferation of cancer, or the evolution of antibiotic drug resistance. However, the sequence samples provided by genomic databases, like the multiple-sequence alignments of homologous proteins considered here, are typically of limited size, including even in the most favorable situations rarely more than $10^3-10^5$ alignable sequences. Fortunately, such sequence data are increasingly complemented by quantitative mutagenesis experiments, which use experimental high-throughput approaches to quantify the effect of thousands of mutants. While it might be tempting to use these data directly to measure mutational landscapes from experiments, it has to be noted that current experimental techniques miss at least 2-3 orders of magnitude in the number of measurable mutants to actually reconstruct the mutational landscape. 

In such situations, where no single dataset is sufficient for accurate inference, integrative methods like the one proposed here will be of major benefit.

\section*{Methods}

\subsection*{Data}
    
	\subsubsection*{Artificial data} For a system of $N=32$ binary spins, couplings $\mathbf{J}^{0}$ and fields $\mathbf{h}^{0}$ are chosen from a Gaussian distribution with zero mean, and standard deviation $\sim 0.8/\sqrt{N}$ for $J$ and $0.2$ for $h$ (analogous results are obtained for other parameter choices, as long as these correspond to a paramagnetic phase). Dataset $\mathbf{D}_{\text{eq}}$ is created by Markov chain Monte Carlo (MCMC) simulation, resulting in $M=100$ equilibrium configurations. A large number ($\sim 10^5$) of MCMC steps are done between each of those configurations to ensure that they are independent. One of these configurations is chosen at random as the reference sequence ("wild-type"), and the energies of all $N$ configurations differing by a single spin flip from the reference are computed ("single mutants"). Gaussian noise of variance $\Delta_0^2$ can be added to these energies, resulting in dataset $\mathbf{D}_E$.

	\subsubsection*{Biological data}
Detailed information about the analysis of biological data is provided in the \emph{Supporting Information}.

\subsection*{Details of the  inference} 
	For artificial data, Eqs.~(\ref{eq:MaxLikelihood_lambda}) are solved using steepest ascent, updating parameters $\mathbf{J}$ and $\mathbf{h}$ in direction of the gradient. To ensure convergence, we have added an additional $\ell_2$-regularization to the joint likelihood: $\gamma_h(\|\mathbf{h}\|_2)^2+\gamma_J(\|\mathbf{J}\|_2)^2$.  A gradient ascent method has been analogously used for the case $\lambda = 1$. 
	To estimate the gradient, it is necessary to compute single and pair-wise probabilities $p_i(\mathbf{J},\mathbf{h})$ and $p_{ij}(\mathbf{J},\mathbf{h})$. Their exact calculation requires summation over all possible configurations of $N$ spins, which is intractable even for systems of moderate size $N$, so we relied on the following approximation schemes.
    
    \subsubsection*{Mean-field inference}
    In the analysis of artificial data we relied on the  mean-field approximation (MF) leading to closed equations for $p_i$ and $p_{ij}$ : 
	\begin{eqnarray}\label{eq:MF_equations}
		&& p_i = e^{ h_i + \sum_{j\neq i}J_{ij}p_j} / (    1+ e^{ h_i + \sum_{j\neq i}J_{ij}p_j} ) \nonumber \\
		&& p_{ij} -p_ip_j = -(\mathbf{J}^{-1})_{ij} \ .
	\end{eqnarray}
	 The main advantage of the MF approximation is its computational efficiency: The first term is solved by an iterative procedure, the second requires the inversion of the couplings matrix $\mathbf{J}$. However, the approximation is only valid and accurate at "high temperatures", \emph{i.e.} small couplings \cite{plefka1982convergence}. This condition is verified in the case of the artificial data described above.

    \subsubsection*{MCMC inference}
     Correlations observed in MSA of protein sequences are typically too strong for the MF approximation to accurately estimate marginals of the model. Therefore we use MCMC sampling of $M^{MC}=10^4$ independent equilibrium configurations to estimate marginals at each iteration of the previously described learning protocol.

\subsection*{Global evaluation of the inferred Ising model}

The Kullback-Leibler divergence $D_{\scriptscriptstyle KL}(P||Q)=\sum_{\vect{s}}P(\vect{s}) \log\{
	P(\vect{s}) / Q(\vect{s}) \}$ is a measure of the difference between probability distributions $P$ and $Q$.  It is zero for $P\equiv Q$, and otherwise positive. 
	In the case of Boltzmann distributions $P\propto e^{-\mathcal{H}_P}$ and $Q\propto e^{-\mathcal{H}_Q}$, its expression simplifies to 
	\begin{equation}\label{eq:kl_boltzmann}
	D_{\scriptscriptstyle KL}(P||Q) = \langle \mathcal{H}_Q-\mathcal{H}_P\rangle_{P} + \log\mathcal{Z}_Q - \log\mathcal{Z}_P. 
	\end{equation}	
    Evaluating this expression requires the exponential computation of the partition function of both models $\mathcal{H}_P$ and $\mathcal{H}_Q$. To overcome this difficulty, we use the symmetrized expression in Eq.~(\ref{eq:kl}), which only involves the average of macroscopic observables. \\
	The symmetrized Kullback-Leibler divergence is computed by obtaining $M_K = 128\,000$ equilibrium configurations from both $P$ and $Q$, using them to estimate the averages in Eq.~(\ref{eq:kl}).


\section*{Acknowledgements}

MW acknowledges funding by the ANR project COEVSTAT (ANR-13-BS04- 0012-01). This work undertaken partially in the framework of CALSIMLAB, supported by the grant ANR-11-LABX-0037-01 as part of the "Investissements d’Avenir" program (ANR-11-IDEX-0004-02).

\section*{Author contributions statement}

M.F. and M.W. designed research; P.B.C., M.F. and M.W. performed research; P.B.C. and M.F.
analyzed data; and P.B.C., M.F. and M.W. wrote the paper.

\section*{Additional information}

The author(s) declare no competing financial interests.

\end{document}


\maketitle

\subsection*{Details of the biological data}

The following datasets are used in the analysis of biological data:
\begin{itemize}
\item {\bf Multiple sequence alignements} We used multiple sequence alignments (MSA) of sequences belonging to the Pfam Beta-lactamase2 family PF13354 for the inference of TEM-1 mutational landscape (resp. PDZ domain family PF00595 for PSD-95 mutational landscape) [30].
Raw data have been processed as described in [18]:
we have used HMMer [29] to search  against the Uniprot protein sequence database (version of March 2015); from the  resulting alignments we have removed all sequences with more than 5 gapped positions; in order to take into account phylogenetic correlations and sampling biases we have associated to each sequence $\vect{s^m}$ the following statistical weight:

\begin{eqnarray}
w_m=\left( 1+ \sum_{m\neq m'}\Theta\Big(d_{m,m'}-\vartheta N\Big) \right)^{-1} \ ,
\end{eqnarray}
with $d_{mm'}$ being the Manhattan distance (number of mismatches) between sequences $\vect{s^m}$ and $\vect{s^{m'}}$ and $\Theta$ being the Heaviside step function whose value is zero for negative argument and one for positive argument. The reweighting threshold is set to $\vartheta=0.8$ as usually done in DCA inference [32]. The resulting MSA are $N=197$ (resp. $N=81$) sites long, and contain 
$M=2462$ (resp. $M=3287$) homologous sequences  (average sequence identity $\sim 20\%$ with the wild-type sequences). 

\item {\bf Computational predictions of thermodynamic stabilities}  We used $\Delta\Delta G$-predictions  for all single amino acid mutations from the wild-type sequences as estimated by PoPMuSic web-based tool [31] taking as reference structure the  PDB: 1M40 for TEM-1  (resp PDB: 1BE9 for PSD-95). 
\item {\bf Experimental fitness measurements} 
\begin{itemize}
\item {\bf TEM-1.} Experimental measurements from [23] provide estimations of amoxicillin Minumum Inibitory Concentration MIC for $n=742$ distinct single amino-acid mutations falling  inside the part of the sequence covered by the corresponding Pfam model. Following [18] we have defined the experimental fitness as the logarithmic average over all MIC measurements of a specific mutant sequence (whenever multiple measurements for the same sequence were available).

\item {\bf PSD-95.} In  [22]  the effect of each single point mutations has been quantified as the log frequency of observing each amino acid $\beta$ at each position $i$ in the selected (sel) versus the unselected (unsel) population, relative to amino acid $s_i$ present in the wild type:

\begin{equation}
\phi(s_i\to \beta)=\log\left[\frac{f_i^{sel}(\beta)}{f_i^{sel}(s_i)}\right]-\log\left[\frac{f_i^{unsel}(\beta)}{f_i^{unsel}(s_i)}\right]
\end{equation}

\end{itemize}
\end{itemize}

\subsection*{Potts Hamiltonian}

The relevant description in the statistical modeling of protein sequences is a 21-state Potts Model, since each variable $s_i, i=1,...,N,$ can now assume 21 states (20 amino acids, one alignment gap). The following Potts Hamiltonian provides the energy of any amino-acid sequence $\vect{s}=(s_1,...,s_N)$,
\begin{equation}
\mathcal{H}^{\rm Potts}(\vect{s})=-\sum_{i<j}^NJ_{ij}(s_i,s_j) -\sum_{i=1}^N h_i(s_i)\ ,
\end{equation}
where fields $h_i$ and couplings $J_{ij}$ are now respectively $21$-dimensional vectors and $21\times21$-dimensional matrices. We fix the gauge freedom (cf.~[1]) using, for all $i,j$, the lattice-gas conditions $h_i(s^{ref}_i)=0$ and 
$J_{ij}(s^{ref}_i,s_j)=J_{ij}(s_i,s^{ref}_j)=0$. Here $\vect{s}^{ref}=(s_1^{ref},...,s_N^{ref})$ is the reference wildtype sequence, for which mutations shall be predicted.

\subsection*{Fitness-energy mapping}

Since model energies $\mathcal{H}(\vect{\sigma^a)}$ can be in any non-linear relation with measured fitnesses $\phi^a$ we replace Eq.~(9) with:
\begin{eqnarray}\label{eq:MaxLikelihood_lambda_mod}
p_{i,\alpha}(\mathbf{J},\mathbf{h}) &=& \pi_{i,\alpha}^{\text{emp}} + \frac{\lambda}{1-\lambda}\frac{1}{M}\sum_{a=1}^P\delta_{\sigma_i^a,\alpha}\left[\phi^a - \mu\big(\mathcal{H}(\vect{\sigma}^a)\big)\right] \\
p_{ij,\alpha\beta}(\mathbf{J},\mathbf{h}) &=& \pi_{ij,\alpha\beta}^{\text{emp}} + \frac{\lambda}{1-\lambda}\frac{1}{M}\sum_{a=1}^P\delta_{\sigma_i^a,\alpha}\delta_{\sigma_j^a,\beta}\left[\phi^a - \mu \big(\mathcal{H}(\vect{\sigma}^a)\big)\right] \nonumber
\end{eqnarray} 
where $p_{i,\alpha}(\mathbf{J},\mathbf{h})$ (resp. $\pi_{i,\alpha}^{\text{emp}}$ ) are the single  marginals in the model (resp. in the empirical sample) specyfing the frequency of observing amino-acid $\alpha$ at position $i$, $p_{ij,\alpha\beta}(\mathbf{J},\mathbf{h})$ (resp. $\pi_{ij,\alpha\beta}^{\text{emp}}$ ) are the pairwise marginals, and
$\mu(x)$ is the robust monotonous non-linear mapping from energy to fitness introduced in [18]. The mapping is obtained by first sorting the model energies and experimental fitnesses and then associating the $n_{th}$ smallest energy $\mathcal{H}(n_{th})$ with the $n_{th}$ highest experimental fitness value $\phi(n_{th})$,
\begin{equation}
\mu\big(\mathcal{H}(n_{th})\big)=\phi(n_{th})\ .
\end{equation}

Note that the second terms in Eq.~(\ref{eq:MaxLikelihood_lambda_mod}) vanish when model energies are exactly in the inverse order of experimental fitnesses. We subsequently used the mapping to compute linear correlations between the mapped energies $\mu\big(\mathcal{H}(\vect{\sigma}^a)\big)$ and the experimental fitnesses $\phi^a$, resulting in non-linear rank correlations between experimental fitnesses and model energies.